%% file: main.tex
\PassOptionsToPackage{hyphens}{url}

\documentclass[sigconf,10pt,authorversion]{acmart}

\usepackage[utf8]{inputenc}
\usepackage{blindtext}
\usepackage{booktabs} 
\usepackage{tabulary}
\usepackage[english]{babel}
\usepackage{graphicx}
\usepackage{subcaption}
\usepackage{balance}
\usepackage{pifont}
\usepackage{multirow}
\usepackage{footnote}
\usepackage{paralist}
\usepackage{algorithm}
\usepackage{algorithmic}
\usepackage{listings}
\definecolor{highlight}{RGB}{192,80,77}
\usepackage{amsmath}

\usepackage[capitalise]{cleveref}
\usepackage[acronym]{glossaries}

\usepackage{bytefield}

\usepackage{ifthen}
\newboolean{extended}
\setboolean{extended}{false}
\newcommand{\extended}[2]{\ifthenelse{\boolean{extended}}{#1}{#2}}

\newcommand{\para}[1]{\smallskip\noindent\textbf{#1.}}

\newcommand{\code}[1]{\texttt{#1}}

\copyrightyear{2018}
\acmYear{2018}
\setcopyright{acmcopyright}
\acmConference[MobiCom '18]{The 24th Annual International Conference on Mobile Computing and Networking}{October 29--November 2, 2018}{New Delhi, India}
\acmBooktitle{The 24th Annual International Conference on Mobile Computing and Networking (MobiCom '18), October 29--November 2, 2018, New Delhi, India}
\acmPrice{15.00}
\acmDOI{10.1145/3241539.3241566}
\acmISBN{978-1-4503-5903-0/18/10}

\begin{document}

\title[The Apple Wireless Direct Link Ad hoc Protocol]{One Billion Apples' Secret Sauce: Recipe for the \\ \emph{Apple Wireless Direct Link} Ad hoc Protocol}

\author{Milan Stute}
\orcid{0000-0003-4921-8476}
\affiliation[obeypunctuation=true]{
	\department{Secure Mobile Networking Lab}\\
	\institution{TU Darmstadt},
	\country{Germany}
}
\email{mstute@seemoo.de}

\author{David Kreitschmann}
\affiliation[obeypunctuation=true]{
	\department{Secure Mobile Networking Lab}\\
	\institution{TU Darmstadt},
	\country{Germany}
}
\email{dkreitschmann@seemoo.de}

\author{Matthias Hollick}
\affiliation[obeypunctuation=true]{
	\department{Secure Mobile Networking Lab}\\
	\institution{TU Darmstadt},
	\country{Germany}
}
\email{mhollick@seemoo.de}

\input{frontback/names}
\input{frontback/glossary}

\begin{abstract}
\input{frontback/abstract}
\end{abstract}

\ccsdesc[500]{Networks~Network protocol design}
\ccsdesc[500]{Networks~Ad hoc networks}
\ccsdesc[500]{Networks~Link-layer protocols}

\keywords{AWDL, Reverse engineering, Ad hoc networks, IEEE\,802.11, Proprietary protocol, Apple, macOS, iOS}

\maketitle

\input{chapter/introduction}
\input{chapter/background}
\input{chapter/methodology}
\input{chapter/overview}
\input{chapter/frames}
\input{chapter/operation}
\input{chapter/evaluation}

\input{chapter/discussion}
\input{chapter/conclusion}

\begin{acks}
This work is funded by the LOEWE initiative (Hesse, Germany) within the NICER project and by the German Federal Ministry of Education and Research (BMBF) and the State of Hesse within CRISP-DA.
\end{acks}

\bibliographystyle{ACM-Reference-Format}
\balance
\bibliography{bibexport,anonymous}

\end{document}

%% file: frontback/names.tex

\newcommand{\timepoint}[1]{\ensuremath{T_\text{#1}}}
\newcommand{\timediff}[1]{\ensuremath{t_\text{#1}}}

\newcommand{\TimePHY}{\timepoint{Tx,PHY}}
\newcommand{\TimeTarget}{\timepoint{Tx,Target}}
\newcommand{\TimeAtNextAW}{\timepoint{AW}}
\newcommand{\TimeToNextAW}{\timediff{AW}}
\newcommand{\TimeDelay}{\timediff{Tx}}

\newcommand{\TimeAir}{\timediff{air}}
\newcommand{\TimeReceive}{\timepoint{Rx}}

\newcommand{\SyncErr}{\ensuremath{\xi}}

\newcommand{\channel}{\ensuremath{C}}
\newcommand{\CurrentChannel}{\ensuremath{C}}

\newcommand{\SeqAW}{\ensuremath{i}}
\newcommand{\ChanSeqLength}{\ensuremath{c}}
\newcommand{\ChanStepCount}{\ensuremath{step}}
\newcommand{\PresenceMode}{\ensuremath{p}}

\newcommand{\Slave}{\ensuremath{\mathcal{S}}}
\newcommand{\Master}{\ensuremath{\mathcal{M}}}

%% file: frontback/glossary.tex

\newacronym{AWDL}{AWDL}{Apple Wireless Direct Link}
\newacronym{PSF}{PSF}{Periodic Synchronization Frame}
\newacronym{MIF}{MIF}{Master Indication Frame}
\newacronym{UMI}{UMI}{Unicast Master Indication Frame}
\newacronym{AW}{AW}{Availability Window}
\newacronym{EW}{EW}{Extension Window}
\newacronym{EAW}{EAW}{Extended Availability Window}
\newacronym{TLV}{TLV}{Type-Length-Value}
\newacronym{TU}{TU}{Time Unit}
\newacronym{AF}{AF}{Action Frame}
\newacronym{mDNS}{mDNS}{Multicast DNS}
\newacronym{BLE}{BLE}{Bluetooth Low Energy}
\newacronym{OUI}{OUI}{Organizational Unique Identifier}
\newacronym{AP}{AP}{Access Point}
\newacronym{NAN}{NAN}{Neighbor Aware Networking}
\newacronym{GO}{GO}{Group Owner}
\newacronym{TDLS}{TDLS}{Tunneled Direct Link Setup}

%% file: frontback/abstract.tex

\gls{AWDL} is a proprietary and undocumented IEEE\,802.11-based ad hoc protocol.
Apple first introduced \gls{AWDL} around 2014 and has since integrated it into its entire product line, including iPhone and Mac.
While we have found that \gls{AWDL} drives popular applications such as AirPlay and AirDrop on more than one billion end-user devices, neither the protocol itself nor potential security and Wi-Fi coexistence issues have been studied.
%
In this paper, we present the operation of the protocol as the result of binary and runtime analysis.
In short, each \gls{AWDL} node announces a sequence of \glspl{AW} indicating its readiness to communicate with other \gls{AWDL} nodes. An elected master node synchronizes these sequences.
Outside the \glspl{AW}, nodes can tune their Wi-Fi radio to a different channel to communicate with an access point, or could turn it off to save energy.
Based on our analysis, we conduct experiments to study the master election process, synchronization accuracy, channel hopping dynamics, and achievable throughput. We conduct a preliminary security assessment and publish an open source Wireshark dissector for \gls{AWDL} to nourish future work.

%% file: chapter/introduction.tex

\section{Introduction}

\glsreset{AWDL}
\glsreset{AW}

\gls{AWDL} is a proprietary protocol deployed in about 1.2 billion\footnote{Based on unit sales for iPhone, iPad, and Mac since 2014~\cite{Apple:Financial}.} end-user devices consisting of Apple's main product families such as Mac, iPhone, iPad, Apple Watch, and Apple TV---effectively all recent Apple devices containing a Wi-Fi chip.
Apple does not advertise the protocol but only vaguely refers to it as a ``peer-to-peer Wi-Fi'' technology~\cite{Apple:NSNetServiceRef,Apple:iOSSecurity}. Yet, it empowers popular applications such as AirDrop and AirPlay that transparently use \gls{AWDL} without the user noticing.
We believe that public knowledge of this undocumented protocol would be beneficial for the following reasons:
First, since \gls{AWDL} is based on IEEE\,802.11, there are potential performance and co-existence issues that need to be identified. This is especially important in regulated environments as \gls{AWDL} uses various channels and employs a channel hopping mechanism that might interfere with corporate Wi-Fi deployments.
Second, the Wi-Fi driver (where \gls{AWDL} is implemented) is the largest binary kernel extension in current versions of macOS. Given the recently published vulnerabilities in Wi-Fi chip firmware~\cite{Beniamini2017:BroadcomPt1,Artenstein2017:Broadpwn} that might lead to full system compromise~\cite{Beniamini2017:BroadcomPt2}, we highly recommend a security audit of the protocol and its implementations as vulnerabilities in non-standardized protocols are even more likely to occur.
For example, protocol fuzzing requires knowledge of the frame format.
Third, an open re-implementation of the protocol would allow interoperability with other operating systems, eventually enabling high-throughput cross-platform direct communication. Such technology is required, for example, in smartphone-based emergency communication applications~\cite{Lu2016:Networking,Kohnhauser2017:SEDCOS}.

%
To maximize the impact for the research community, we have lifted a layer in Apple's ecosystem and unveiled an existing yet obscure wireless ad hoc protocol.
In this paper, we conduct a comprehensive investigation on \gls{AWDL} by means of binary and runtime analysis, and present its frame format and operation.
In short, \gls{AWDL} is based on the IEEE\,802.11 standard and makes use of vendor-specific extensions that allow custom protocol implementations.
Each \gls{AWDL} node periodically emits custom action frames containing a sequence of \glspl{AW} indicating its readiness to communicate with other \gls{AWDL} nodes. An elected master node synchronizes these sequences.
Within these \glspl{AW}, nodes are able to communicate with their neighbors using a dedicated data frame format.
Outside the \glspl{AW}, nodes can tune their Wi-Fi radio to a different channel to communicate with an access point, or turn it off to save energy.
We summarize our main contributions:
\begin{itemize}
	\item We provide insights into the macOS operating system and its Wi-Fi driver architecture and debugging facilities to help future research endeavors (\cref{sec:methodology}).
	\item We present the \gls{AWDL} frame format and operation in detail (\cref{sec:overview,sec:frames,sec:operation}).
	\item We conduct an experimental analysis of \gls{AWDL} to assess election behavior, synchronization accuracy, throughput, and channel hopping strategies (\cref{sec:experiments}).
	\item We discuss protocol complexity, energy efficiency, and perform a preliminary security assessment where we report a security permission problem in a macOS kernel extension (\cref{sec:discussion}).
	\item We publish an open source \gls{AWDL} Wireshark dissector~\cite{awdl-wireshark}.
\end{itemize}
%
Furthermore, we give background on related direct wireless communication technologies in \cref{sec:background} and conclude this work in \cref{sec:conclusion}.

%% file: chapter/background.tex

\begin{figure}
	\includegraphics[width=0.9\linewidth]{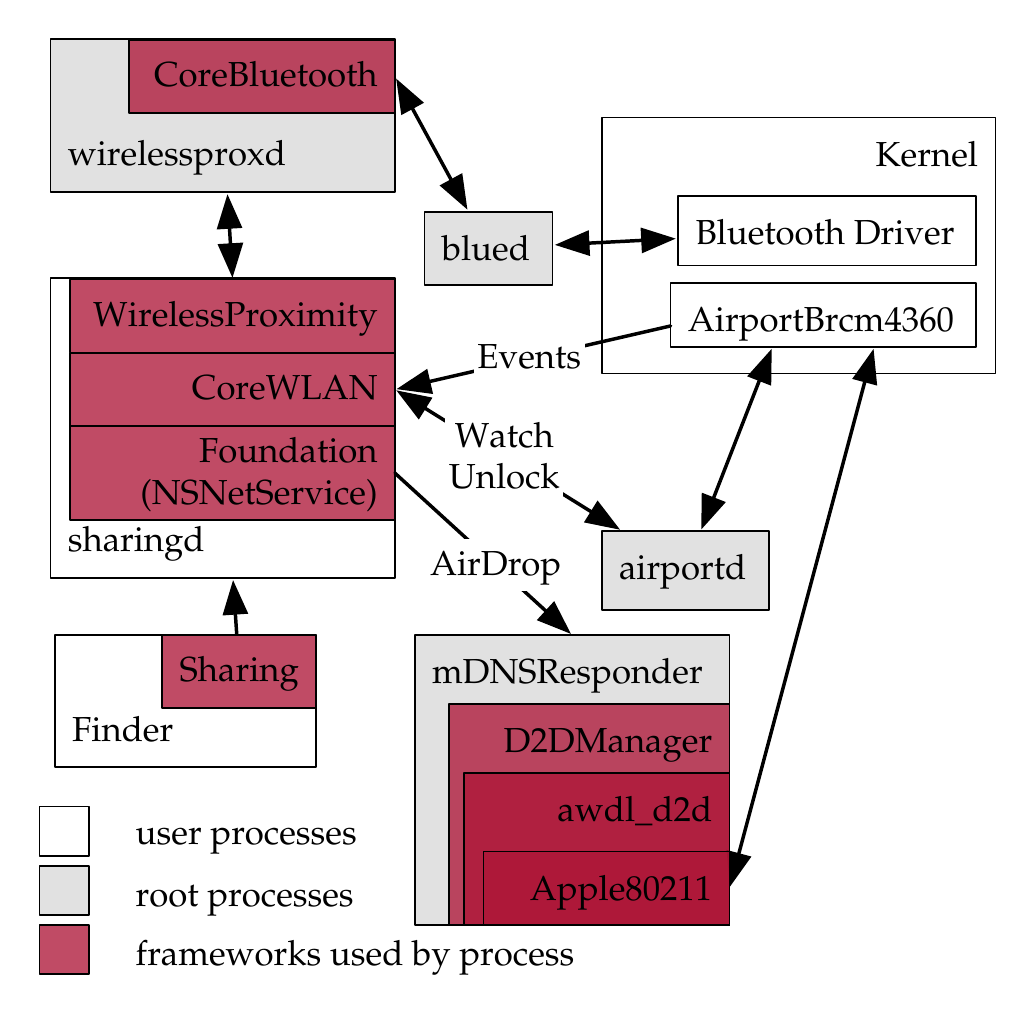}
	\caption{Interaction of different macOS processes and frameworks used for controlling \gls{AWDL}.}
	\label{fig:sharingd}
\end{figure}

\section{Background}
\label{sec:background}

\gls{AWDL} has been referenced in several patents such as~\cite{Linde2016:RTWifi} and can be classified as a wireless ad hoc protocol which allows peers to communicate directly with each other.
There exist already a number of other technologies which we summarize in the following.

\para{IEEE\texorpdfstring{\,}{ }802.11 IBSS}
The IBSS mode commonly known as ``ad hoc'' mode creates a distributed wireless network without special controller roles.
An IBSS is created by sending beacon frames with an SSID and BSSID on a particular channel. Other nodes joining the network will send out beacons themselves using the same information.
The mode is robust to nodes leaving the network as all nodes broadcast beacons. The nodes do not require any further synchronization.
However, IBSS has never become widely deployed, mostly due to lack of efficient power saving mechanisms, which are crucial for mobile devices \cite{CampsMur2013:WiFiDirect}. Flawed implementations are another common problem \cite{ServalProject}. IBSS is not supported on Android and Microsoft announced it might not be available in future versions of Windows \citep{Microsoft:AdhocAPI}.
On Apple's operating systems encryption is not supported and iOS only allows to join existing IBSS networks.

\para{Wi-Fi Peer-to-Peer}
Wi-Fi P2P \citep{WifiAlliance2016:P2P}, also known under its certification name Wi-Fi Direct\footnote{The Wi-Fi Alliance is a vendor association which holds the Wi-Fi trademark for IEEE\,802.11-based technologies and certifies products using the specification. Although the Wi-Fi Alliance does not formally create the standard, their certification has relevance in the market. The alliance also creates their own standards based on IEEE\,802.11 such as P2P and NAN.}, allows connecting multiple devices directly without a base station.
During operation, one node assumes the role of a \gls{GO} which closely resembles infrastructure (or BSS) operation.
It is not possible to migrate the role of the \gls{GO} to another device: if the \gls{GO} leaves the network, a new network must be created.
Wi-Fi P2P connections are established by listening on one channel and sending probe requests on all channels. This delays the connection process in practice. Experiments show that establishing a connection takes from four to more than ten seconds~\cite{CampsMur2013:WiFiDirect}. Discovering devices thus drains their battery very fast.


\para{Tunneled Direct Link Setup}
\gls{TDLS} is an IEEE\,802.11 extension that enables direct communication between two nodes in the same BSS. In networks without \gls{TDLS}, all traffic passes the \gls{AP} even when the two communicating nodes are within communication range. \gls{TDLS} requires both nodes to be connected to the same \gls{AP} since control frames are tunneled through the \gls{AP} and, thus, cannot be used in real ad hoc scenarios.

\para{Neighbor Awareness Networking}
\gls{NAN}~\cite{WifiAlliance2015:NAN}, also known as Wi-Fi Aware, extends IEEE\,802.11 with proximity service discovery. \gls{NAN} is designed to be energy efficient, allowing continuous operation on battery-powered devices~\cite{CampsMur2015:NAN}.
\gls{NAN} is supported in Android 8~\cite{Google2017:AndroidNAN}, but we did not find any devices with compatible hardware.
\gls{NAN} depends on beacon frames sent from an elected master. These synchronize the timing of all devices in an area. During a short discovery window the master sets, devices can turn their radio on, exchange service and connection information (e.g., parameters for Wi-Fi P2P) and turn their radio off again.
In fact, we found that \gls{AWDL} employs similar concepts as \gls{NAN}, but the actual implementation differs strongly from that of \gls{NAN}. In addition, \gls{NAN} does not feature a data path for transmission of user data.

\para{Bluetooth}
Bluetooth~\cite{Bluetooth2016:CoreSpec} is a separate standard with different PHY and MAC layers.
Bluetooth operates in the 2.4~GHz band as Wi-Fi and is often integrated into the Wi-Fi chip to share the same antennas.
%
\gls{BLE} is incompatible with classic Bluetooth and is optimized for low energy consumption and, therefore, offers limited bandwidth.
The usable maximum \gls{BLE} 4.2 data rate is 394~kbit/s \cite{Dooley2017:AppleBluetooth}. It is commonly implemented in small battery-powered devices such as smartwatches and fitness trackers.
\gls{BLE} is not designed for large data transfers but can be used for bootstrapping high-bandwidth links such as \gls{AWDL}.

%% file: chapter/methodology.tex

\section{Methodology}
\label{sec:methodology}

Reverse engineering is more of an art than a science and, hence, it is hard to write generic recipes. Nevertheless, we structure our methodology for reversing closed-source network protocols with a focus on the macOS operating system so that it can be used in related research endeavors.
In the following, we describe how binary and dynamic runtime analysis in tandem can result in full disclosure of the workings of a complex wireless network protocol.
Previous exemplary works have reverse engineered the Skype protocol~\cite{Ouanilo:SkypeReverse}, Broadcom Wi-Fi chip firmware~\cite{Schulz2018:Nexmon}, and the Fitbit ecosystem~\cite{Classen2018:Fitbit}.

\subsection{Binary Analysis}
\label{sec:binary-analysis}

\extended{
\begin{table}
	\caption{Largest kernel extensions in macOS 10.12.6 according to the output of \code{kextstat}.}
	\label{tab:kext-size}
	\begin{tabular}{@{} lr @{}}
		\toprule
		\textsc{Kernel Extension} & \textsc{Size (KB)} \\
		\midrule
		\textbf{com.apple.iokit.IO80211Family} & \textbf{920} \\
		com.apple.iokit.IOThunderboltFamily & 936 \\
		com.apple.driver.DspFuncLib & 1308 \\
		\{\dots\}.driver.AppleIntelBDWGraphicsFramebuffer & 1504 \\
		\textbf{com.apple.driver.AirPort.Brcm4360} & \textbf{7924} \\
		\bottomrule
	\end{tabular}
\end{table}
}{}

We analyzed numerous binaries related to \gls{AWDL} to finally find those parts that implement the protocol. We first illustrate our selection process and then discuss the two-part Wi-Fi driver which implements most of the \gls{AWDL} protocol stack. We focus our analysis on macOS and assume that the architecture is in principle similar to that of iOS. We used a decompiler to analyze the target binaries.

\para{Binary Selection}
Apple excessively uses \emph{frameworks} and \emph{daemons} in its OSes. Consequently, there are numerous dependencies which result in a complex binary selection process. Frameworks offer an API to their corresponding singleton daemons and can be used by other daemons and processes.
We started off by crawling the system for binaries that had ``802.11'', ``\gls{mDNS},'' or ``sharing'' in their names. We found more related targets by following dependencies.
We show part of the discovered dependencies and interactions in \cref{fig:sharingd}.
While there are user-facing binaries such as the \code{sharingd} daemon, the most relevant binaries reside in the kernel, in particular, the generic Wi-Fi driver \code{IO80211Family} and the device-specific variants \code{AirportBrcm4360} and \code{Airport\-BrcmNIC}. Each of them includes hundreds of AWDL-related functions, suggesting that the bulk of the protocol stack is implemented here.
We found that \code{IO80211Family} takes care of most of the \gls{AWDL} frame parsing and creation as well as maintaining the \gls{AWDL} state machine. The device-specific driver handles time-critical functions such as synchronization.
As both driver parts are among the largest kernel extensions present in macOS \extended{(see \cref{tab:kext-size})}{}, understanding internal driver structures were key to make sense of the decompiled code.

\para{Finding Interesting Code Segments}
Due to the size of the macOS Wi-Fi driver, we needed to quickly find functions that would implement part of the \gls{AWDL} protocol.
Fortunately, Apple does not strip symbol names from their binaries, such that searching for ``awdl'' in the symbol table (e.\,g., using \code{nm}) results in a number of hits. Some of those symbols additionally contain ``parse'' and ``TLV'' in their name (e.\,g. \code{parseAwdlSyncTreeTLV}) which helped us understand the calculation of some \gls{TLV} fields.
Furthermore, debug log statements give hints about the purpose of a code segment inside a function. Therefore, we can search for debugging strings and their cross-references to find details such as the misalignment threshold in \cref{sec:synchronization}.

\para{Leaked Broadcom Driver Source Code}
%
%
%
As another source of information, we used a dated Broadcom Wi-Fi driver whose source code was leaked \cite{GitHub:LeakedBCMDriver}.
We found several references to \gls{AWDL} in the source code but none of the core functionality.
We suspect that Broadcom uses a modular firmware concept with one central repository for a wide range of features. Special features such as \gls{AWDL} are made available selectively to their customers such as Apple.
%
More important than the references to \gls{AWDL} are some C structs found in the source code. These include key structures such as the Synchronization Parameters TLV and Channel Sequence TLV (more in \cref{sec:frames}).
The leaked code also contains the source code for the \code{wl} utility, which provides debugging features for the driver and is further discussed in \cref{sec:method:dynamic}.

\begin{figure}
	\includegraphics[width=1\linewidth]{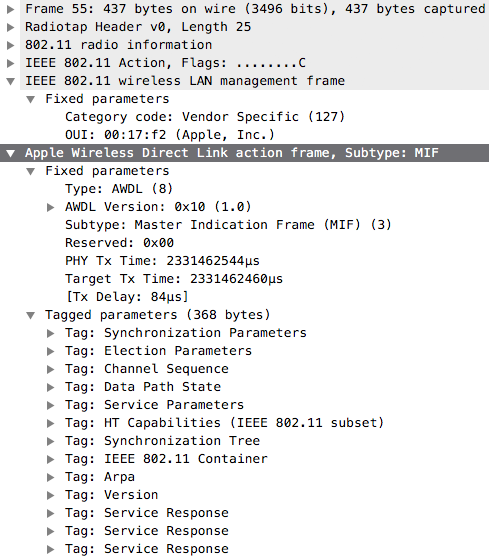}
	\caption{Screenshot of our Wireshark dissector.}
 	\label{fig:wireshark}
\end{figure}

\para{Dissecting Structures}
To understand the driver's functions, we needed to reconstruct the underlying data structures. The leaked source code shows that most of the \gls{AWDL}-related functions use an \code{awdl\_info} struct as a first parameter. The \code{wlc\_dump\_awdl} function prints internal data in a readable format and, thus, was an ideal target to reconstruct the internal structures as shown below:
\small
\begin{lstlisting}[language=C,morekeywords={bcm_bprintf,bcm_ether_ntoa}]
bcm_bprintf(a2, "AWDL master home channel = %d\n",
             awdl_info->master_home_channel);
\end{lstlisting}
\normalsize
%
%
%
The result of our binary analysis was a complete Wireshark dissector for \gls{AWDL} that we also used for the dynamic analysis of the protocol and for evaluating our experiments.
We show our dissector in \cref{fig:wireshark}.

\subsection{Runtime Analysis}
\label{sec:method:dynamic}

The complete protocol operation was difficult to comprehend with the binary analysis alone. To understand the semantics of synchronization, election, service discovery, and data path, we complemented our static analysis with a dynamic approach.
In this section, we discuss dedicated macOS logging and debugging facilities that helped to analyze the protocol.
In particular, we used the \emph{Console} application, the \code{ioctl} interface, the leaked Broadcom \code{wl} utility, as well as Apple's undocumented \emph{CoreCapture} framework. The latter is especially verbose but required us to write an additional dissector for Wireshark as it uses a private data format.

\para{Apple Console}
The \emph{Console} program is the central place to access logs since macOS 10.12 and includes debug messages from the kernel.
To receive verbose output from the Wi-Fi driver, we increased the log level using custom boot arguments which we found by searching for references to the \code{PE\_parse\_boot\_arg} function in the Wi-Fi driver.
The following boot arguments maximize the driver's debug output:
\small
\begin{lstlisting}[language=bash,keywords={nvram}]
nvram boot-args="debug=0x10000 \
       awdl_log_flags=0xffffffffffffffff \
       awdl_log_flags_verbose=0xffffffffffffffff \
       awdl_log_flags_config=1 wlan.debug.enable=0xff"
\end{lstlisting}
\normalsize
With the increased log level, \emph{Console} shows additional information such as state transitions and the current channel sequence:
\small
\begin{lstlisting}
IO80211Family <...> com.apple.p2p: AWDL ON: [infra(100) 72%], (6/44/44) [44 0 0 0 0 0 0 0 6 44 44 0 0 0 0 0] Low Power
\end{lstlisting}
\normalsize

\para{\code{ioctl} Interface}
\code{ioctl} system calls are a standard way to communicate with devices on Unix-based systems. Apple uses \code{ioctl}s to configure wireless interfaces such as associating with an \gls{AP} or creating an IBSS.
Apple provides the header files with the request format, the available request types, and the data structures for macOS 10.5. These old header files can be brought up to date using information from the binary analysis. The \code{apple80211VirtualRequest} method contains calls to all handler functions. Out of the available 72 request IDs, 40 relate to \gls{AWDL}. These requests can set several parameters in the driver.
Especially useful is the card-specific \code{ioctl}. It allows wrapping a Broadcom-specific \code{ioctl} inside an Apple \code{ioctl}, providing us with a direct interface with the Broadcom driver.
Note that it is no longer possible to send Broadcom-specific \code{ioctl}s since Apple fixed our reported vulnerability (\cref{sec:discussion}): the driver now checks for a private \emph{entitlement} security permissions~\cite{Apple:Entitlements} (\code{com.apple.driver.AirPort.Broadcom.ioctl\-access}) which requires a binary signed with an Apple private key.
It should be possible to overwrite the respective permission-checking function in the driver using a kernel extension patching framework such as~\cite{GitHub:Lilu} to restore unrestricted \code{ioctl} access. Driver patching requires disabling Apple's \emph{System Integrity Protection}~\cite{Apple:SIP}.

\para{Broadcom \code{wl} Utility}
The Broadcom \code{wl} utility found in the leaked source code provides several methods to access internal information about \gls{AWDL} operations, which are directly related to the structures found during binary analysis.
Although the \gls{AWDL}-specific driver code was missing in the leaked source code, the \code{wl} source code contains \gls{AWDL} related commands and structures.
\code{wl} allows us to query the current \gls{AWDL} driver status using commands such as \code{dump awdl} and \code{awdl\_advertisers}. The latter shows information about neighboring nodes including RSSI.

\para{CoreCapture Framework}
CoreCapture is Apple's primary logging and tracing framework for IEEE\,802.11 on iOS and macOS. CoreCapture combines raw protocol traces with traditional log entries and provides snapshots of the device and driver state.
CoreCapture is undocumented but was referenced in a \code{dumpPacket} function that we found in the driver. Since the framework outputs (among other logs and memory dumps) numerous PCAP trace files with a custom header format, we wrote a Wireshark dissector for CoreCapture that we make available to the public~\cite{awdl-wireshark}.
In addition, we publish a manual for CoreCapture with this paper~\cite{corecapture-manual}.

%% file: chapter/overview.tex
\section{AWDL Overview}
\label{sec:overview}

\begin{figure}
	\includegraphics[width=0.9\linewidth]{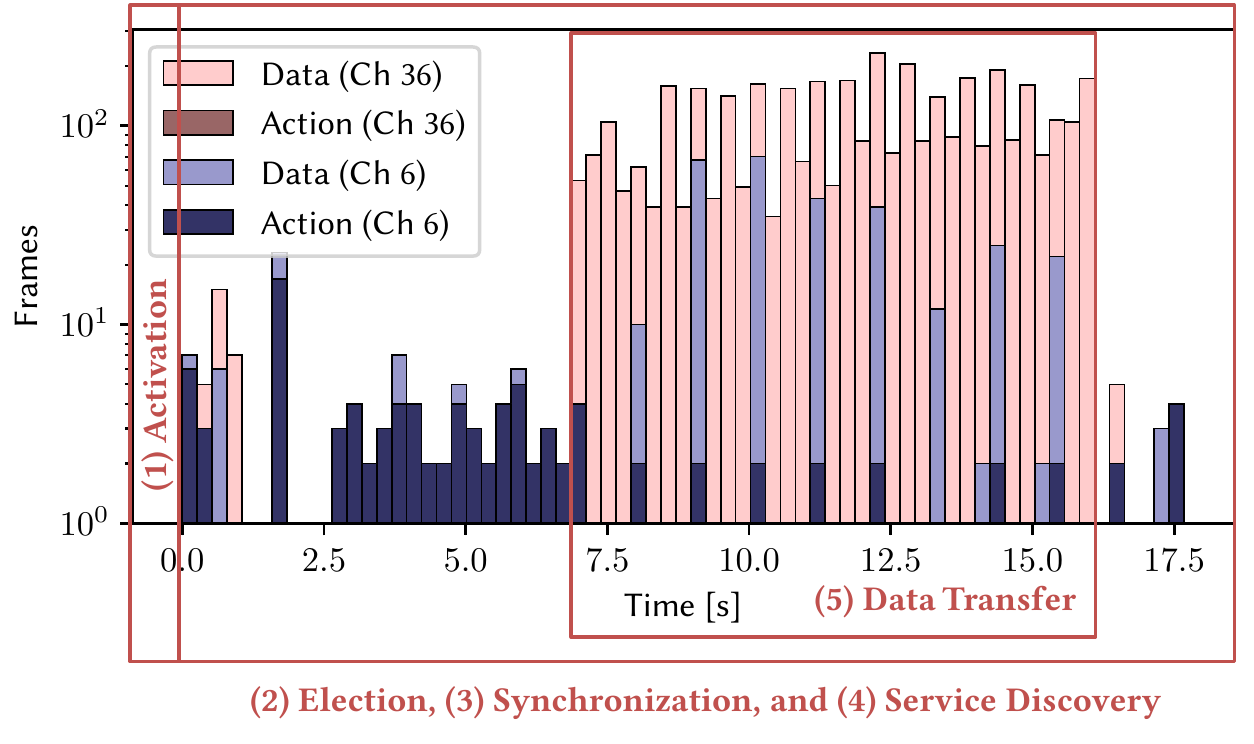}
	\caption{The five main \gls{AWDL} phases. Exemplary trace showing a 100\,MiB file transfer via AirDrop including activity on multiple channels and differentiating the traffic types.}
	\label{fig:example-airdrop}
\end{figure}

Based on our analysis, we formulate \emph{hypotheses} regarding the design goals and decisions of \gls{AWDL}:
\begin{inparaenum}[($i$)]
	\item leverage existing hardware (Wi-Fi chip), thus building the protocol on top of IEEE\,802.11;
	\item conserve energy, especially on mobile devices, hence synchronizing and putting the Wi-Fi chip into a power-saving mode during idle times;
	\item allow seamless operation of direct and infrastructure-based communication, so enable synchronized channel hopping without disconnecting from an AP; and
	\item enable fast service discovery, thus offloading DNS-SD to Wi-Fi frames.
\end{inparaenum}
Commodity Wi-Fi chips usually have a single RF chain and are, therefore, restricted to a single wireless channel at any given time. To use multiple channels, an adapter needs to switch channels and cannot use the regular wireless connection for short periods of time. This is expected behavior for roaming (scan for available networks while being connected to a network) and power saving features (switch off the radio).
To use these short periods for data transfer, devices need a method for discovery and coordination when to meet on which channel.
We depict the main \gls{AWDL} phases in \cref{fig:example-airdrop} and briefly introduce them in the following.

\para{(1) Activation}
Apple uses \gls{AWDL} as an on-demand communication technology. This means that \gls{AWDL} is inactive by default, but applications can (temporarily) request activation.
For example,
\begin{inparaenum}[]
	\item AirDrop uses \gls{BLE} for activation by sending truncated hashes of the user's contact information;
	\item AirPlay receivers (Apple TV) constantly announce their presence via \gls{AWDL}; and
	\item third-party application may activate the interface indirectly by advertising services via the \code{NSNetService} API~\cite{Apple:NSNetServiceRef}.
\end{inparaenum}

\para{(2) Master Election}
Apple uses fixed social channels (6, 44, and 149\footnote{depending on the country}) for coordination using \glspl{PSF}. A node starting its \gls{AWDL} interface monitors the social channels for some time to discover other nodes in range. If \gls{AWDL} \glspl{AF} are received, the node can adopt an existing master. If no frames are received, it assumes the master role itself.
We elaborate on the election process in \cref{sec:election}.

\para{(3) Synchronized Channel Sequences}
\glsreset{AW}
\glsreset{EAW}
\gls{AWDL} is built around a sequence of time slots (\glspl{AW} and \glspl{EAW}). For each of these slots, peers broadcast if they are available for \gls{AWDL} data and, if so, on which channel they will be. Peers match these advertisements with their own \gls{AW} sequence. If there is a common channel in a particular \gls{AW}, communication during this \gls{AW} is possible. A synchronization mechanism aligns the sequences between nodes.
We elaborate on the synchronization and channel alignment processes in \cref{sec:synchronization,sec:chanseq}, respectively.

\para{(4) Service Discovery}
DNS Service Discovery (DNS-SD)~\cite{RFC2782} also known as ``Bonjour'' can be offloaded to \gls{AWDL}. \gls{AWDL} piggy-backs DNS-SD responses directly onto its \glspl{AF} such that services are immediately discovered whenever a node changes its advertisements. For space reasons, we do not elaborate on the service discovery component in this paper.

\para{(5) Data Transfer}
\gls{AWDL} uses a vendor-specific frame format header for user data which exclusively transports IPv6 packets.
When transmitting user data to a particular peer, a node needs to calculate the \glspl{AW} during which both nodes are tuned to the same channel and only transmit frames during those \glspl{AW}.
In addition, \gls{AWDL} adapts its channel sequence according to the current outgoing traffic load.
We discuss the data transfer mechanisms in detail during the experimental evaluation in \cref{sec:experiments}.

%% file: chapter/frames.tex

\section{Frame Format}
\label{sec:frames}

\glsreset{PSF}
\glsreset{MIF}

We discovered two general frame types used by \gls{AWDL}: \emph{action} and \emph{data} frames which are used for coordination and direct data transfer, respectively. We elaborate on the frame format of these types in the following.

\subsection{Action Frames}

\begin{figure}
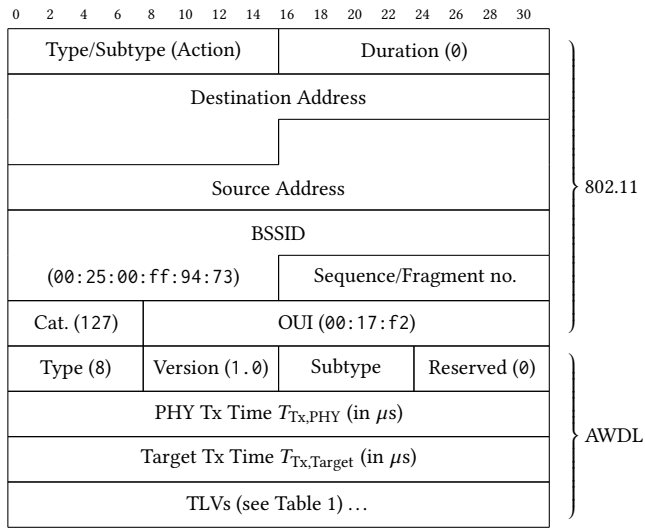

\centering
\footnotesize
\begin{bytefield}[bitwidth=0.8em]{32}
	\bitheader{0,2,4,6,8,10,12,14,16,18,20,22,24,26,28,30} \\
	\begin{rightwordgroup}{802.11}
		\bitbox{16}{Type/Subtype (Action)} & \bitbox{16}{Duration (\code{0})} \\
		\wordbox[tlr]{1}{Destination Address} \\
		\bitbox[blr]{16}{} & \bitbox[tlr]{16}{} \\
		\wordbox[blr]{1}{Source Address} \\
		\wordbox[tlr]{1}{BSSID} \\
		\bitbox[blr]{16}{(\code{00:25:00:ff:94:73})} & \bitbox{16}{Sequence/Fragment no.} \\
		\bitbox{8}{Cat.\ (\code{127})} & \bitbox{24}{OUI (\code{00:17:f2})}
	\end{rightwordgroup} \\
	\begin{rightwordgroup}{AWDL}
		\bitbox{8}{Type (\code{8})} & \bitbox{8}{Version (\code{1.0})} & \bitbox{8}{Subtype} & \bitbox{8}{Reserved (\code{0})} \\
		\wordbox{1}{PHY Tx Time \TimePHY{} (in $\mu$s)} \\
		\wordbox{1}{Target Tx Time \TimeTarget{} (in $\mu$s)} \\
		\wordbox{1}{TLVs (see \cref{tab:tlvs}) \dots}
	\end{rightwordgroup}
\end{bytefield}
\caption{\gls{AWDL} action frame.}
\label{fig:action-frame}
\end{figure}

\gls{AWDL} uses IEEE\,802.11 vendor-specific \glspl{AF} which generally allow vendors with an \gls{OUI} to implement IEEE\,802.11 frames with arbitrary payloads~\cite{IEEE:80211-2016}. The \gls{AWDL} vendor-specific extension consists of a fixed-sized header and multiple \gls{TLV} fields as shown in \cref{fig:action-frame}.
A \gls{TLV} consist of a 1-byte \emph{type} field, followed by a 2-byte \emph{length} field which indicates the length of the subsequent \emph{value} byte string.
%
The fixed header mostly includes static values such as \gls{AWDL}-specific BSSID, OUI, version, and type.
The two timestamps indicate when the frame was created and, therefore, at which time the included information was up-to-date (\TimeTarget{}), and when it was actually queued for transmission (\TimePHY{}). Their difference approximates the sender's transmission delay and is used for synchronization purposes.
There are two \gls{AWDL} \gls{AF} subtypes: \gls{PSF} and \gls{MIF}.
These frame types start with the same fixed header and differ only in the included set of \glspl{TLV} and, hence, their size.
We show the frame format excluding the FCS at the end of the frame in \cref{fig:action-frame}.
We first explain the purpose of the subtypes and then discuss \glspl{TLV} used in \gls{AWDL}.


\glsreset{PSF}
\para{\gls{PSF}}
The \gls{PSF} is used for synchronization and is further explained in Chapter \ref{sec:synchronization}. The name was gathered from a patent~\cite{Vandwalle2016:PatentAWDL}. Its subtype is \code{0}.
If all participating devices support the 5 GHz band, the \gls{PSF}
is the only frame type also seen on the 2.4 GHz band.

\glsreset{MIF}
\para{Broadcast \gls{MIF}}
The \gls{MIF} is used for multiple purposes, e.g., election (\cref{sec:election}) and service discovery. It includes more \glspl{TLV} and is sent by all devices in the network regularly. The \gls{MIF} subtype is \code{3}.


\para{\glspl{TLV}}
\glspl{TLV} contain the actual control information. The different types can be attributed to one of the following purposes: \emph{election and synchronization}, \emph{service discovery}, and \emph{user data transmission}. In addition, the \emph{version} TLV provides a 1-byte version number which presumably supersedes the version field in the fixed header (see \cref{fig:action-frame}).
We summarize all \glspl{TLV} in \cref{tab:tlvs} and discuss them briefly in the following. The names were taken from function names and debugging strings found during binary analysis.
We discuss only some \glspl{TLV} in detail in this paper, and refer to our Wireshark dissector for the full specification.
Note that some \emph{type} values (e.\,g. 1, 3, and 8) are missing in \cref{tab:tlvs}. These types appear to be deprecated as they were not actively used in the \gls{AWDL} versions that we analyzed.

\begin{table}
\small
\newcommand{\yes}{\ding{51}}
\newcommand{\multiple}{+}
\caption{\glspl{TLV} used in \gls{AWDL}. \textnormal{We give the name and type value of a \gls{TLV}, indicate whether it is included in \glspl{PSF} or \glspl{MIF} (\yes), and whether it can be present multiple times (\multiple).}}
\label{tab:tlvs}
\begin{tabular}{@{} lcccl @{}}
	\toprule
	\textsc{Name} & \textsc{Type} & \textsc{\gls{PSF}} & \textsc{\gls{MIF}} & \textsc{Purpose} \\
	\midrule
	Sync.\ Parameters & 4 & \yes & \yes & \multirow{5}{2.1cm}{Election and Synchronization} \\
	Channel Sequence & 18 & \yes & \yes & \\
	Election Parameters & 5 & \yes & \yes & \\
	Election Parameters v2 & 24 & \yes & \yes & \\
	Synchronization Tree & 20 & \yes & \yes & \\
	\midrule
	Service Parameters & 6 & \yes & \yes & \multirow{3}{2.1cm}{Service Discovery} \\
	Service Response & 2 & & \yes\multiple & \\
	Arpa (Reverse DNS) & 16 & & \yes & \\
	\midrule
	Data Path State & 12 & \yes & \yes & \multirow{3}{2.1cm}{User Data Transmission} \\
	HT Capabilities & 7 & & \yes & \\
	VHT Capabilities & 17 & & \yes & \\
	\midrule
	Version & 21 & \yes & \yes & Compatibility \\
	\bottomrule
\end{tabular}
\end{table}

The \emph{election and synchronization} processes handle the overall cooperation of the devices. The data in these \glspl{TLV} determines, e.\,g., which node takes the master role and which channels are to be used.
Curiously, the Synchronization Parameters \gls{TLV} includes its own channel sequence, so the separate Channel Sequence \gls{TLV} appears to be redundant. It was however always transmitted on current operating system versions.
This is further discussed in \cref{sec:election,sec:synchronization,sec:chanseq}.
The \emph{service discovery} components offload \gls{mDNS} and DNS-SD functionality to the \glspl{AF}. They contain the hostname (Arpa \gls{TLV}); and PTR, SRV, and TXT resource records (Service Response \gls{TLV}).
The \emph{user data transmission} components are used to negotiate the parameters for direct connections between devices. For example, supported PHY rates are announced in the HT/VHT Capabilities \glspl{TLV} which are similar to the ones introduced in the IEEE\,802.11\,n and 11\,ac amendments~\cite{IEEE:80211-2016}.
%
In addition, each peer announces in the Data Path State \gls{TLV} the Wi-Fi network (BSSID) that it is currently connected to as well as the real MAC address of the Wi-Fi chip. We believe that this information could be used to offload an \gls{AWDL} connection to an infrastructure network if both peers are connected to the same network. However, this would require additional reachability tests due to network policies such as client isolation, and we did not observe such behavior in practice.
The \emph{version} \gls{TLV} includes the \gls{AWDL} version (half a byte for major and minor version number each) as well as a device class ID. We found that v3.x is used in macOS~10.13 and iOS~11; and v2.x in macOS~10.12 and iOS~10 (and potentially prior iOS versions). AWDL v1.x is used in macOS~10.11 which does not support the \emph{version} \gls{TLV}. The device class seems to indicate the OS type of the node, e.\,g., macOS (1) or iOS (2).

\subsection{Data Frames}

\begin{figure}
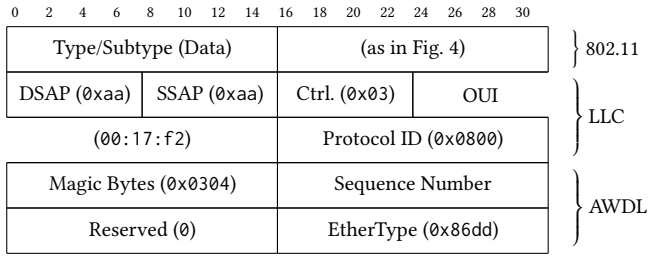

	\centering
	\footnotesize
	\begin{bytefield}[bitwidth=0.8em]{32}
		\bitheader{0,2,4,6,8,10,12,14,16,18,20,22,24,26,28,30} \\
		\begin{rightwordgroup}{802.11}
			\bitbox{16}{Type/Subtype (Data)} & \bitbox{16}{(as in \cref{fig:action-frame})}
		\end{rightwordgroup} \\
		\begin{rightwordgroup}{LLC}
			\bitbox{8}{DSAP (\code{0xaa})} & \bitbox{8}{SSAP (\code{0xaa})} & \bitbox{8}{Ctrl.\ (\code{0x03})} & \bitbox[tbl]{8}{OUI} \\
			\bitbox[tbr]{16}{(\code{00:17:f2})} & \bitbox{16}{Protocol ID (\code{0x0800})}
		\end{rightwordgroup} \\
		\begin{rightwordgroup}{AWDL}
			\bitbox{16}{Magic Bytes (\code{0x0304})} & \bitbox{16}{Sequence Number} \\
			\bitbox{16}{Reserved (\code{0})} & \bitbox{16}{EtherType (\code{0x86dd})}
		\end{rightwordgroup}
		\extended{\\
		\begin{rightwordgroup}{IPv6}
			\wordbox{2}{\dots}
		\end{rightwordgroup}
		}{}
	\end{bytefield}
	\caption{\gls{AWDL} data frame header.}
	\label{fig:data-frame}
\end{figure}

AWDL uses IEEE\,802.11 data frames for user data transmission. The To-DS and From-DS flags are set to zero, similar to IBSS which means that these frames are addressed directly, and three address fields are used for the destination, source, and BSSID. We depict the \gls{AWDL} data frame format in \cref{fig:data-frame}.
The BSSID in AWDL frames is always \code{00:25:00:ff:94:73} which belongs to the \gls{OUI} \code{00:25:00} that is assigned to Apple~\cite{IEEE:Registry}.
%
%
The LLC header contains a different Apple \gls{OUI} (\code{00:17:f2}) and a protocol ID in the SNAP part. These headers are part of the IEEE\,802 standard \citep{IEEE:802-2014} and allow vendors to implement their own protocols on higher layers.
%
%
The actual \gls{AWDL} data header essentially consists of a sequence number and the EtherType of the transported protocol. We identified IPv6 as the only protocol used with \gls{AWDL}.

\subsection{Addressing for Higher-Layer Protocols}

\gls{AWDL} is used in conjunction with higher-layer protocols. Therefore, it needs some way to address \gls{AWDL} nodes via a network layer protocol.
This is especially important because \gls{AWDL} implements privacy-enhancing MAC randomization which means that instead of using the Wi-Fi chip's fixed MAC address, it generates a random address every time the interface is activated.
In IPv6, address resolution is usually done via the Neighbor Discovery Protocol (NDP).
Apple, however, does not use NDP for \gls{AWDL}, but instead generates link-local IPv6 addresses from the source address field contained in the \glspl{AF} (\cref{fig:action-frame}) using a method described in RFC\,4291~\cite[Appendix A]{RFC4291}.
This method constructs a link-local IPv6 address based on the 48-bit MAC address of the network interface.
In particular, given a 48-bit MAC address \code{o\textsubscript{0}:o\textsubscript{1}:o\textsubscript{2}:o\textsubscript{3}:o\textsubscript{4}:o\textsubscript{5}}, the corresponding link-local IPv6 address is constructed as:
\begin{lstlisting}{morekeywords={80}}
fe:80::(*@\textbf{o\textsubscript{0}}@*)^0x02:(*@\textbf{o\textsubscript{1}}@*):(*@\textbf{o\textsubscript{2}}@*):ff:fe:(*@\textbf{o\textsubscript{3}}@*):(*@\textbf{o\textsubscript{4}}@*):(*@\textbf{o\textsubscript{5}}@*).
\end{lstlisting}
%
Using this standardized method, nodes can add their neighbors to the neighbor table immediately after receiving the first \gls{AF} without the need or overhead of an additional address resolution protocol such as NDP or ARP.

%% file: chapter/operation.tex

\section{Protocol Operation}
\label{sec:operation}

We present the detailed mechanisms that are used to form and maintain an \gls{AWDL} \emph{cluster}. In particular, we discuss how a master is elected, and conflicts are resolved; how nodes synchronize their clock to the master; and, finally, how the announced channel sequence maps to the sequence of \glspl{AW}.

\subsection{Master Election}
\label{sec:election}

In this section, we explain the election process and the tree-based synchronization structure. In particular, we focus on the mechanisms that make \gls{AWDL} robust to master nodes leaving or joining the cluster.

\para{Role of the Master Node}
As already mentioned, \gls{AWDL} relies on roughly synchronous clocks of all participating nodes in a cluster.
To achieve this, it is paramount that there is exactly one node in the cluster which has the responsibility of emitting a ``clock signal.''
This is the one (and as far as we know the only) role of the \emph{master} node.
All other nodes in the cluster are called \emph{slaves} and should adopt this signal.
In a simple scenario with only two nodes, one node will be the master and another a slave.
In larger scenarios, slave nodes might be more than one hop away from the master node. In such cases, intermediate slave nodes will take the role of \emph{non-election masters}, which have the responsibility to repeat the master's clock signal.
The intermediate master nodes are included in the Synchronization Tree \gls{TLV} where each node announces the path to the ``top'' master. In any case, there is only one top master in a cluster.

\para{Master Metric}
The master election is based on a \emph{metric} field which is included in the Election Parameters v2 \gls{TLV}.
The node that announces the largest metric value will become the master of that cluster.
Apple's patent \cite{Vandwalle2016:SyncPatent} claims that these metrics could be based on available energy resources, CPU load, signal strength, etc.
In practice, however, the metric is simply chosen at random.
A node that activates its \gls{AWDL} interface initially sets its metric field to 60 and listens on the social channels for an existing master for 2 seconds. If no master is found, it draws a random number from a predefined range and sets this as its metric.
We have found that this range depends on the \gls{AWDL} version, e.\,g., 405 to 436 in v2.x and 505 to 536 in v3.x. We assume that this is done for backwards compatibility so that the master node is guaranteed to be a node running the most up-to-date version in a cluster and future protocol extensions can be supported.

\para{Merging Clusters with Different Masters}
When two already established \gls{AWDL} clusters with different master nodes move into proximity, they need to merge such that nodes in the different clusters will be able to discover each other.
In \gls{AWDL}, the process is straight-forward as all nodes advertise their current master metric in the Election Parameters \gls{TLV}.
If two nodes with different masters discover each other, they receive the top master metric of the other cluster and can immediately adopt the master with the higher metric. The remaining nodes in the ``lower'' cluster then follow as soon as the first node advertises the new master metric. 




\para{Loop Prevention}
When creating such an election tree hierarchy with multiple levels of sub-masters, loops may occur.
To prevent loops and limit the maximum depth of the election tree, each \gls{AF} contains a list of all nodes up to the top master in the Synchronization Tree TLV.
Each node can then make sure that it does not adopt a non-election master if it is already in that node's path.

\begin{figure}
\centering
\footnotesize
\begin{bytefield}[bitwidth=0.9em]{32}
	\bitheader{0,2,4,6,8,10,12,14,16,18,20,22,24,26,28,30} \\

	\bitbox{8}{Type (\code{4})} & \bitbox{16}{Length} & \bitbox{8}{TX Channel} \\
	\bitbox{16}{Tx Counter \TimeToNextAW} & \bitbox{8}{Master Channel} & \bitbox{8}{Guard Time (\code{0})} \\
	\bitbox{16}{AW Period (\code{16})} & \bitbox{16}{AF Period (mostly \code{110} or \code{440})} \\
	\bitbox{16}{Flags} & \bitbox{16}{AW Extension Length (\code{16})} \\
	\bitbox{16}{AW Common Length (\code{16})} & \bitbox{16}{Remaining AW} \\
	\bitbox{8}{Ext.\ Min (\code{3})} & \bitbox{8}{Multicast Max.\ (\code{3})} & \bitbox{8}{Unicast Max.\ (\code{3})} & \bitbox{8}{AF Max.\ (\code{3})} \\
	\bitbox[tlr]{32}{Master MAC Address} \\
	\bitbox[blr]{16}{} & \bitbox{8}{Presence Mode (\code{4})} & \bitbox{8}{Reserved (\code{0})} \\
	\bitbox{16}{Sequence Number \SeqAW{}} & \bitbox{16}{AP Beacon Alignment} \\
	\wordbox{1}{Channel Sequence (as in \cref{fig:chanseq})}
\end{bytefield}
\caption{Synchronization Parameters \gls{TLV}}
\label{fig:syncparams}
\end{figure}

\para{Re-Election}
The initialization of a device using a low metric will prevent most random re-elections when new devices join the network. As there is no sign-off message, a master leaving the network simply stops sending \glspl{AF}. Therefore, a missing master can only be detected by other devices after a certain \emph{no master} timeout which is fixed to 96 \glspl{AW} ($\approx$ 1.5\,s).
Another node will then take the place of the old master. As this node was already in sync with the old master, other slave nodes do not need to re-synchronize but simply adopt the new master.
In other words, \gls{AWDL} is robust to ``master churn,'' i.\,e., a leaving master does not interrupt communication, and a new master is seamlessly adopted. This is in contrast to other technologies such as Wi-Fi Direct, where the respective master node essentially acts as an \gls{AP} which takes care of relaying data between two nodes and a leaving master would require a group re-establishment.

\para{The Role of RSSI}
The RSSI values of received \glspl{AF} are used to filter out possibly unstable connections.
In particular, \gls{AWDL} nodes drop frames when the RSSI is below a so-called \emph{edge sync} threshold which is set to -65 (or -78 if AirPlay is used).
Frames from the current master node are accepted with a lower RSSI. These frames receive a bonus \emph{slave sync} threshold of 5.
Lowering the threshold for the master frames allows for a certain variance in the RSSI. We assume that this was done to reduce ``master flapping'' where a node frequently adopts a new master because it regularly drops frames and the \emph{no master} timeout occurs.


\subsection{Synchronization}
\label{sec:synchronization}

\glsreset{AW}

\begin{figure}
\centering
\includegraphics[width=0.9\linewidth]{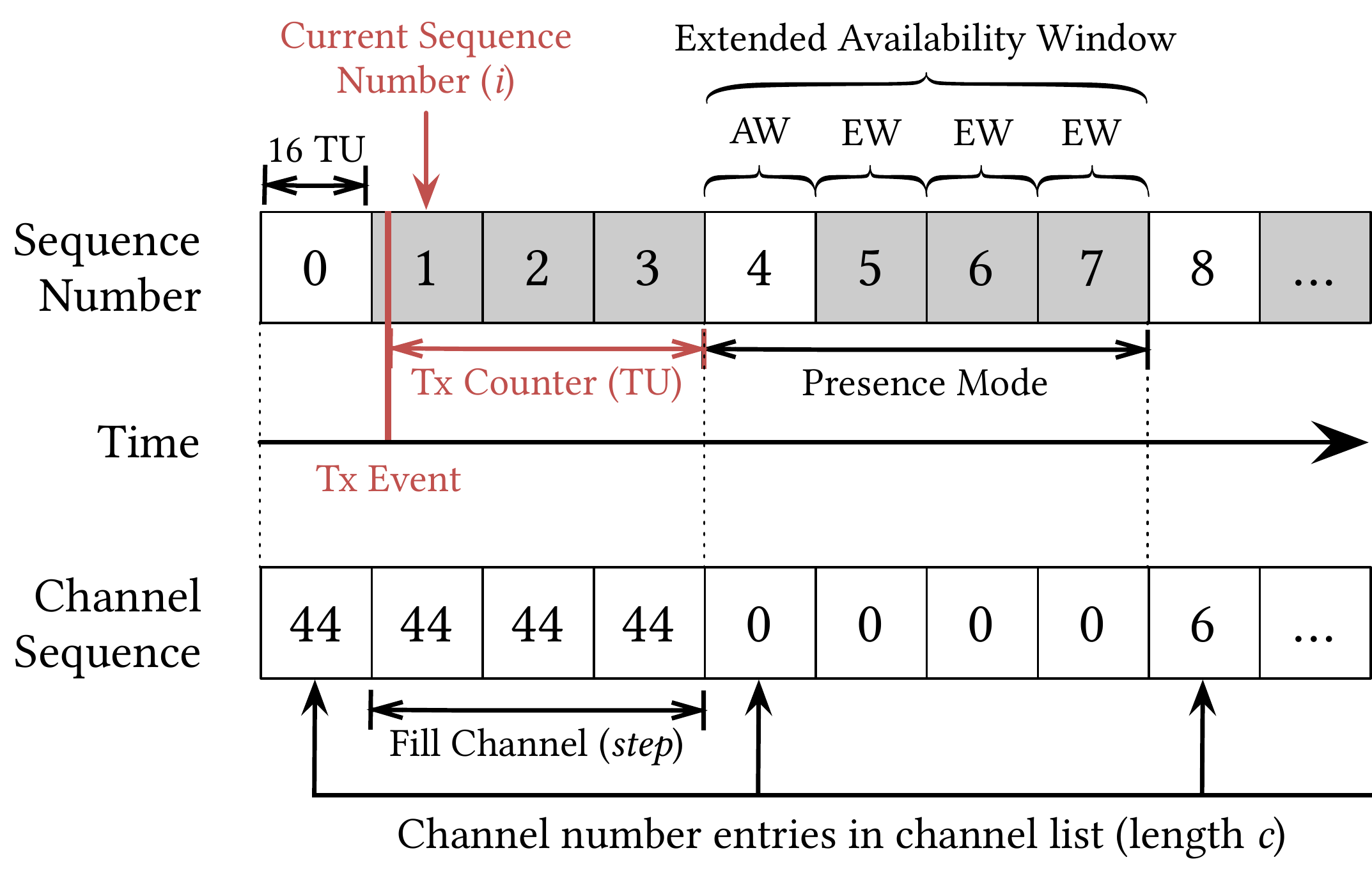}
\caption{Structure of \glspl{AW} and mapping to channel sequence.}
\label{fig:sync-sequence}
\end{figure}

Synchronization is tightly coupled with the election process since nodes always try to synchronize to their elected master.
In this section, we describe how time is structured in \gls{AWDL} and how nodes align their time reference with that of their master.
We introduce the concept of \glspl{AW}, that is, short fixed-length time slots during which communication is possible. These windows have a static length, but can be extended using \glspl{EW}. Finally, we show how the start of an \gls{AW} is determined using fields from the Synchronization Parameters TLV.
We summarize the key concepts and variables in \cref{fig:sync-sequence}.

\para{Availability Window}
\glspl{AW} indicate a period of time during which a device will be available for communication. These windows need to be synchronous for all nodes in a cluster such that every device starts an \gls{AW} at the same time.
Timing in \gls{AWDL} is based on \glspl{TU} where $1\,\mathrm{TU}=1024\,\text{\textmu s}$ \citep[page 141]{IEEE:80211-2016}. In the \gls{AWDL} implementation, an \gls{AW} is always set to be 16 \glspl{TU} long. The length of an \gls{AW} and all other ``static'' values presented in this section are contained in the Synchronization Parameters TLV.
In theory, different configurations are possible, but we found that only fixed values are used.
%

\para{Presence Mode and Extension Windows}
\glsreset{EW}
\glsreset{EAW}
For reduced power consumption, a peer can indicate that it is not listening in every \gls{AW}. A presence mode \PresenceMode{} of \code{4}, which is the only value used in Apple's AWDL implementation, means that a peer is only listening for every fourth window.
%
If a node is transmitting or receiving data, it may extend its time spent on the channel. This is called an \gls{EW}. A presence mode of \code{4} leaves space for three \glspl{EW} of 16 \glspl{TU}.
In addition, \gls{AWDL} allows to configure different numbers of unicast, multicast, and \gls{AF} \glspl{EW}, but these fields are currently always set to 3 and, thus, align with the presence mode.
\Cref{fig:syncparams} shows the parameters transmitted in the Synchronization Parameters TLV.
Given the static configuration, the effective smallest time unit in use is four consecutive \glspl{AW}/\glspl{EW}. For the remainder of this paper, we use the term \gls{EAW} to refer to such a 64\,\gls{TU} time slot.

\para{Calculating the Start of an Availability Window}
Each slave node needs to synchronize its clock to that of its master node.
To achieve this, the master node announces the \emph{start of the next \gls{AW}}.
When transmitting an \gls{AF}, the master includes the number of \glspl{TU} to the next \gls{EAW} \TimeToNextAW{} as well as the sequence number of the current \gls{AW} or \gls{EW} \SeqAW{}.
We mark these values in red in \cref{fig:sync-sequence}.

As these values are set when the frame is created in the driver, some time passes until the frame is actually transmitted via the Wi-Fi interface.
\gls{AWDL} tries to compensate for this transmitter delay by including two additional timestamps in the fixed header of each \gls{AF}: the PHY and target transmission times \TimePHY{} and \TimeTarget{}, respectively.
Ideally, \TimeTarget{} is set when the frame is created, and \TimePHY{} just before the frame is transmitted via the interface. In the macOS driver, however, both timestamps are set in the Wi-Fi driver and, therefore, do not account for delays induced by the distributed coordination function (DCF) which controls medium access~\cite{IEEE:80211-2016}.
%
%
Nevertheless, a device receiving an \gls{AF} from its master at time \TimeReceive{} can approximate the start of the next \gls{AW} \TimeAtNextAW{} as follows:
\begin{align}
	\TimeAtNextAW = \TimeToNextAW \cdot 1024 - (\TimePHY - \TimeTarget) + \TimeAir + \TimeReceive. \label{eq:sync}
\end{align}
In fact, \gls{AWDL} ignores the airtime \TimeAir{} since it is in the order of sub-$\mu$s in a typical close-range Wi-Fi scenario, and the accepted synchronization error is 3\,ms.\footnote{In the function \code{awdl\_recv\_action\_frame}, a misalign metric is increased if the difference between a projection from a previous calculation and new calculation of \TimeAtNextAW{} is larger than 3\,ms.} We experimentally evaluate the achievable accuracy in \cref{sec:experiments}.

\subsection{Channel Sequence}
\label{sec:chanseq}

The \gls{AWDL} channel sequence announcement builds upon the synchronized \glspl{AW} and indicates whether a node is actually available for communication and, if so, on which channel it has tuned its radio.
In this section, we explain
\extended{
the IEEE\,802.11 channel hierarchy,
how the individual channels are encoded in \gls{AWDL}, and
}{}
how the channel sequence maps to the sequence of \glspl{AW}.

\begin{figure}
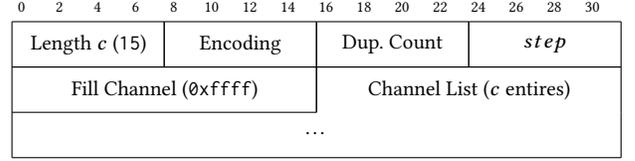

\centering
\footnotesize
\begin{bytefield}[bitwidth=0.9em]{32}
	\bitheader{0,2,4,6,8,10,12,14,16,18,20,22,24,26,28,30} \\

	\bitbox{8}{Length \ChanSeqLength{} (\code{15})} & \bitbox{8}{Encoding} & \bitbox{8}{Dup.\ Count} & \bitbox{8}{\ChanStepCount{}} \\
	\bitbox{16}{Fill Channel (\code{0xffff})} &
	\bitbox[tlr]{16}{Channel List (\ChanSeqLength{} entires)} \\
	\wordbox[blr]{1}{\dots} \\
\end{bytefield}
\caption{Channel Sequence}
\label{fig:chanseq}
\end{figure}

\extended{
\para{IEEE\,802.11 Channel Hierachy}
IEEE\,802.11 channels are defined not only by a channel number, but rather by a bandwidth and by a primary channel.
The primary channel is used for backward compatibility with older devices which only support 20\,MHz channels. Management frames such as beacons or \gls{AWDL} \glspl{AF} are transmitted on the 20\,MHz primary channel so that all devices are able to receive them.
The full bandwidth is then used for unicast frames if both parties have expressed support for HT or VHT data rates.
IEEE\,802.11 groups channels in the 5\,GHz band via a binary tree where the leaf nodes are the 20\,MHz channels and bandwidth doubles on each level towards the tree root.
This way, each 20\,Mhz primary channel can be extended to a wider channel be following the path up to the tree root, thus, yielding a well-defined center frequency for each combination of bandwidth and primary channel.

\para{Channel Encoding}
\gls{AWDL} announces its channel sequence redundantly as part of the Synchronization Parameters TLV as well as in the dedicated Channel Sequence TLV.
The latter appears to be used in recent versions of \gls{AWDL} as the contained channels use the IEEE\,802.11 \emph{operating class} encoding which supports channels with bandwidths above 40\,MHz, thus enabling communication with devices supporting VHT data rates.
In contrast, the \emph{legacy} encoding used in the Synchronization Parameters TLV only supports channels with a maximum bandwidth of 40\,MHz.
We have found a third encoding during binary analysis which simply uses the channel number and is exclusively used by the Apple Watch which only supports the 2.4\,GHz band.

\para{Mapping the Channel Sequence to \glspl{AW}}
}{}
The channel sequence maps channel numbers to \gls{AW} sequence numbers.
While the channel sequence included in the \glspl{TLV} shown in \cref{fig:chanseq} contains a fixed number of $\ChanSeqLength{} + 1 = 16$ channel entries, the sequence can be prolonged with the \ChanStepCount{} field similar to the presence mode, so that one channel entry can span multiple \glspl{AW} and \glspl{EW}.\footnote{Note that the extension with \ChanStepCount{} works only if the \emph{fill channel} field is set to \code{0xffff}, which was the case in all our captured frames.}
Setting \ChanStepCount{} to \code{1} means that the channel will be active for one additional \gls{AW}. However, Apple always sets this field to \code{3}, meaning that the channel will be active for four \glspl{AW} or one \gls{EAW}. Thus, the channel sequence is fully aligned to the presence mode in the Synchronization Parameters \gls{TLV}.
Given an encoded channel sequence and an \gls{AW} sequence number \SeqAW{}, an \gls{AWDL} node can calculate the currently active channel \CurrentChannel{} for any peer based on the following calculation:
\begin{align}
	\CurrentChannel{} &= \SeqAW \mod \left(\left(\ChanSeqLength + 1\right) \cdot \left(\ChanStepCount + 1\right)\right)
\end{align}
As Apple uses fixed values for \ChanSeqLength{} and \ChanStepCount, the announced channel sequence covers  $(15+1) \cdot (3+1) = 64$ \glspl{AW} which takes about one second ($64\,\text{AW} \cdot 16\,\text{TU}/\text{AW} = 1048576\,\text{\textmu{}s} \approx 1\,\text{s}$).

%% file: chapter/evaluation.tex

\section{Experimental Analysis}
\label{sec:experiments}

We analyze the runtime behavior of \gls{AWDL} in different scenarios to
\begin{inparaenum}[($i$)]
	\item validate our findings of the previous sections and
	\item assess the performance of the protocol.
\end{inparaenum}
First, we describe our test setup. Then, we look at the master election and synchronization accuracy in an idle scenario without data transmissions.
We further analyze the channel hopping behavior and throughput performance.

\subsection{Test Setup}
Our test setup consists of one monitoring device and a number of Apple devices.
Our monitor device is an APU board\footnote{\url{https://pcengines.ch/apu.htm}} equipped with two Qualcomm Atheros QCA9882 Wi-Fi cards to support simultaneous sniffing on two different channels which are tuned to \gls{AWDL}'s primary (44) and secondary (6) channel.
Both Wi-Fi cards support hardware timestamping which mitigates variable delays in the receiver's OS stack.
To synchronize the internal clocks of the sniffing Wi-Fi chips, we start each experiment with a calibration phase: we tune both chips to a common channel and let them record multiple frames. Post-experiment, we calculate the timestamp difference of frames that were received by both cards. We use the median difference to correct the clock offset and align both traces.
All following experiments were conducted inside a Faraday tent to minimize interference.
Our test devices include
an iPhone\,8 (iOS~11.2.2),
an iPad\,Pro\,10.5" (iOS~11.0.3),
an iMac (Late 2012, macOS~10.12.6), and
a MacBook Pro (Late~2015, macOS~10.12.6).

\begin{figure}
	\begin{minipage}[c]{\linewidth}
	\centering
	\includegraphics[width=\linewidth]{gfx/plot/idle-7/{{merged.pcapng-master}}}
	\caption{Master selection over time.}
	\label{fig:eval:master}
	\end{minipage}
	\begin{minipage}[c]{\linewidth}
	\centering
	\includegraphics[width=\linewidth]{gfx/plot/idle-7/{{merged.pcapng-selfmetric}}}
	\caption{Self metric over time. \textnormal{The grey shaded areas show the value ranges used in the different versions of \gls{AWDL}.}}
	\label{fig:eval:selfmetric}
	\end{minipage}
\end{figure}

\subsection{Master Election}
\label{sec:eval:election}

In our first experiment, we analyze the master election process.
We observe an \gls{AWDL} cluster in an idle state, meaning that no data transmission takes place and the only observed frames are \glspl{AF}.
We use a setup consisting of an iPhone, iPad, iMac, and MacBook.
We activate the \gls{AWDL} interface by selecting the \emph{sharing panel} in one device which causes a \gls{BLE} scan and activates the \gls{AWDL} interface of other devices in range (on iOS, this only works if the device is unlocked).
To get more interesting results, we let the different devices join approximately 30\,s after one another.

\Cref{fig:eval:master} shows the currently selected master of each node.
First, the iMac creates the \gls{AWDL} cluster and consequently selects itself as the master. As soon as the iPhone joins, it takes over the master role, and the iMac adopts it. The MacBook runs the same version as the iMac and, thus, after having discovered the \gls{AWDL} cluster, it also adopts the iPhone as the master node.
The iPad briefly adopts the existing master, but then immediately takes over this role as it selects a higher self metric than the iPhone:
\cref{fig:eval:selfmetric} shows the current self metric of each node over time. We show the initial value of 60 and the implemented ranges for the different versions of \gls{AWDL}.
Finally, all nodes successively leave the cluster (Wi-Fi turned off) until only the MacBook remains.
Since the iMac and the MacBook run an older version of \gls{AWDL}, they are only selected as master if none of the newer versions are present in the cluster.

Most of these results were expected. What is interesting, however, is that an already existing master node can be ``overtaken'' by another node running the same version of \gls{AWDL}.
This indicates that Apple's \gls{AWDL} implementation is rather simplistic: each node keeps the initial self metric only for a short period of time and then selects a higher random value from the version-dependant range \emph{irrespective of whether it has found an existing master or not}.

\subsection{Synchronization-to-Master Accuracy}
\label{sec:eval:sync}

\begin{figure}
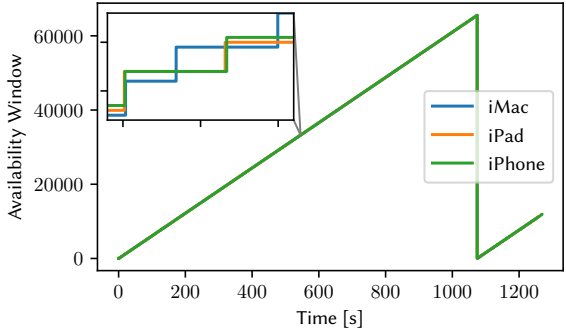

	\includegraphics[width=0.9\linewidth]{gfx/plot/idle-5-20min/{{idle-5.pcapng-sync}}}
	\caption{\gls{AW} sequence number. \textnormal{Showing the sequence number wrap after approximately 18\,min ($\approx 2^{16}\,AW \cdot 16 \frac{TU}{AW} \cdot 1024 \frac{\mu{}s}{TU}$).}}
	\label{fig:aw-seq}
\end{figure}

\begin{figure}
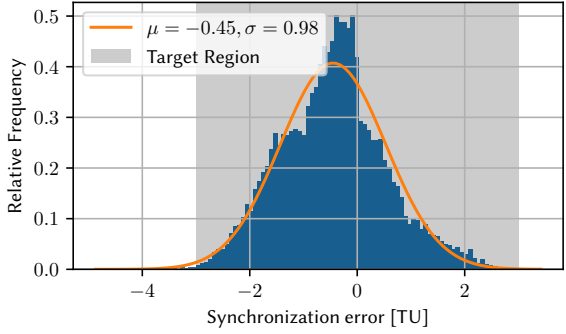

	\includegraphics[width=0.9\linewidth]{gfx/plot/idle-5-20min/{{idle-5.pcapng-syncerr}}}
	\caption{Distribution of synchronization error \SyncErr{}.}
	\label{fig:syncerr}
\end{figure}

We want to evaluate how well AWDL's master election and synchronization mechanism work. To this end, we monitor the \gls{PSF} and \gls{MIF} exchanges between a number of different nodes.
We run another idle experiment over a longer period of time (20 min) with three nodes.
\Cref{fig:aw-seq} shows the \gls{AW} sequence number each node advertises.
While \cref{fig:aw-seq} indicates that synchronization works in principle (all nodes follow the same \gls{AW} sequence number incline), we can see that the \gls{AW} sequence number steps are not perfectly aligned.
We are interested in the magnitude of this synchronization offset.
We adapt \cref{eq:sync} to compute the synchronization error \SyncErr{} between a slave \Slave{} and its master \Master{}. Assuming a constant airtime \TimeAir{} and given two \glspl{AF} from \Slave{} and \Master{} with a sequence number in the same \gls{EAW} recorded at the sniffer at time $\TimeReceive^\Master$ and $\TimeReceive^\Slave$, respectively, we calculate \SyncErr{} as
\begin{align}
\begin{split}
	\SyncErr &= \TimeAtNextAW^\Master - \TimeAtNextAW^\Slave \\
	\extended{
	&= \left(\TimeToNextAW^\Master \cdot 1024 - \left(\TimePHY^\Master - \TimeTarget^\Master\right) + \TimeAir + \TimeReceive^\Master\right) - \\
	&\phantom{{}=} \left(\TimeToNextAW^\Slave \cdot 1024 - \left(\TimePHY^\Slave - \TimeTarget^\Slave\right) + \TimeAir + \TimeReceive^\Slave\right) \\
	}{}
	&= \left(\TimeToNextAW^\Master - \TimeToNextAW^\Slave\right) \cdot 1024 - \left(\TimeDelay^\Master - \TimeDelay^\Slave\right) + \TimeReceive^\Master - \TimeReceive^\Slave, \\
	&\phantom{{}=} \forall i_\Slave{}, i_\Master{} \text{ with } \lfloor \frac{i_\Slave{}}{\PresenceMode{}} \rfloor = \lfloor \frac{i_\Master{}}{\PresenceMode{}} \rfloor.
\end{split}
\end{align}
In \cref{fig:syncerr}, we can see that the synchronization error approximates a Gaussian distribution with a mean value of -0.45, and a standard deviation of 0.98.
\Cref{fig:syncerr} also shows that the target maximum synchronization error of 3\,\glspl{TU} is met in more than 99\,\% of all cases.

While the results are within the target region, the relatively large synchronization error leads to the conclusion that only a portion of each \gls{EAW} can reliably be used for communication and the 3\,\glspl{TU} have to used as a guard interval.
In numbers, this means that only $1 - \frac{2 \cdot 3\,\gls{TU}}{64\,\gls{TU}} \approx 90.6\,\%$ of the interval can be used for communication.
The main source of synchronization error lies in the calculation of the transmission delay \TimeDelay{}. \Cref{eq:sync} assumes that \TimePHY{} is set exactly at the moment when the frame is being transmitted via the Wi-Fi radio after the frame has already been enqueued and additional DCF back-offs have expired. However, we have found that in macOS, \TimePHY{} is set in the driver right after the \gls{AF} is created and before DCF has been run. We did not analyze the implementation for other OSes but assume that this is done at a similar location.

\subsection{Channel Activity}
\label{sec:eval:chan-activity}

\begin{figure}
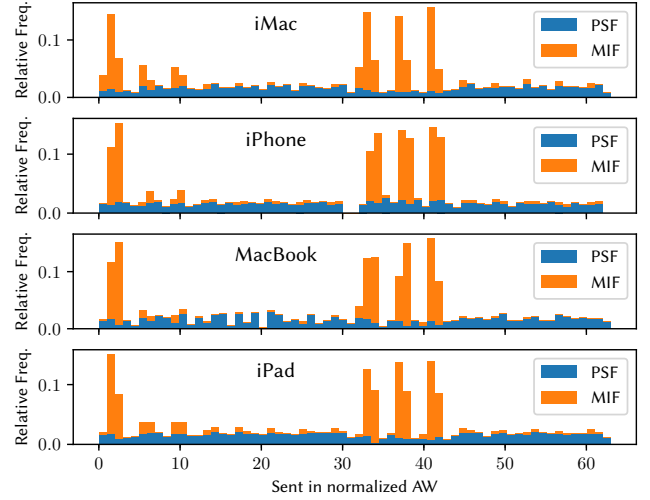

	\centering
	\includegraphics[width=1\linewidth]{gfx/plot/idle-7/{{merged.pcapng-activechan}}}
	\caption{Activity in a full channel sequence period.}
	\label{fig:activechan}
\end{figure}

We want to find out when \glspl{AF} are usually transmitted.
For this, we consider the \emph{idle} scenario from \cref{sec:eval:election} again.
\Cref{fig:activechan} shows when frames (\gls{MIF} and \gls{PSF}) are transmitted during an \gls{EAW} by the different nodes. Each bin represents a single \gls{AW} (16\,TU).
We notice that \glspl{MIF} are mostly sent at the beginning of the first and second half of the entire sequence.
\extended{
We confirm this by looking at the advertised channel list:
\cref{fig:channellist} shows the relative frequency of the different channel numbers in the advertised channel list. We can see that slot 9 is always reserved for channel 6.
}{}
\begin{figure}
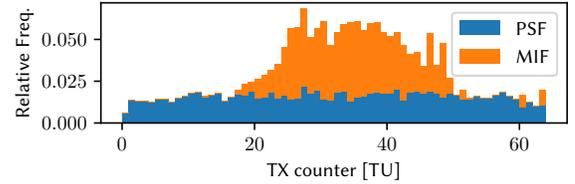

	\centering
	\includegraphics[width=0.9\linewidth]{gfx/plot/idle-7/{{merged.pcapng-txcounter}}}
	\caption{Activity within a single \gls{EAW}.}
	\label{fig:txcounter}
\end{figure}
We also notice that there is a distinct difference in the sending behavior of \glspl{MIF} and \glspl{PSF}.
While \gls{MIF} transmissions adhere to the advertised channel sequence, \glspl{PSF} are sent at any time.
This is probably due to the \gls{AF} period in the Synchronization Parameters \gls{TLV} (see \cref{fig:syncparams}) which is either set to 110 or 440 \gls{TU} and does not align with the 64 \glspl{AW} that cover one channel sequence.
We do not have a solid explanation for this design decision but suspect that it could accelerate the bootstrapping of new nodes which have not yet synchronized to a master node.
As a downside, this means that nodes cannot really go to a power-conserving mode in a non-advertised slot, which we assumed to be one of the core design goals of \gls{AWDL}.
%
Another interesting aspect is that \glspl{PSF} are sent by \emph{all} nodes, no matter if they are master or not. This is another indicator that energy efficiency was not a primary goal of \gls{AWDL}. Otherwise, only the master and sub-masters would send \glspl{PSF}.

\extended{
\begin{figure}
	\includegraphics[width=0.8\linewidth]{gfx/plot/idle-7/{{merged.pcapng-channellist}}}
	\caption{Advertised channel list in \emph{idle} scenario.}
	\label{fig:channellist}
\end{figure}
}{}

\Cref{fig:activechan} also shows that the \glspl{PSF} constitute a certain baseline ``noise,'' while the \glspl{MIF} are sent especially during the middle of one \gls{EAW}.
\Cref{fig:txcounter} ``zooms in'' and depicts the channel activity within a single \glspl{EAW}. We see that \gls{MIF} activity is clustered around the center, while \glspl{PSF} are sent with equal probability over the entire \gls{EAW}.
We think that \glspl{MIF} are considered more important since they contain more information than \glspl{PSF} (see \cref{tab:tlvs}) and sending in the middle of an \gls{EAW} increases the chance that a node receives a transmission even if they are not perfectly synchronized.

\subsection{Throughput and Channel Hopping}
\label{sec:eval:iperf}

\begin{figure}
	\includegraphics[width=0.9\linewidth]{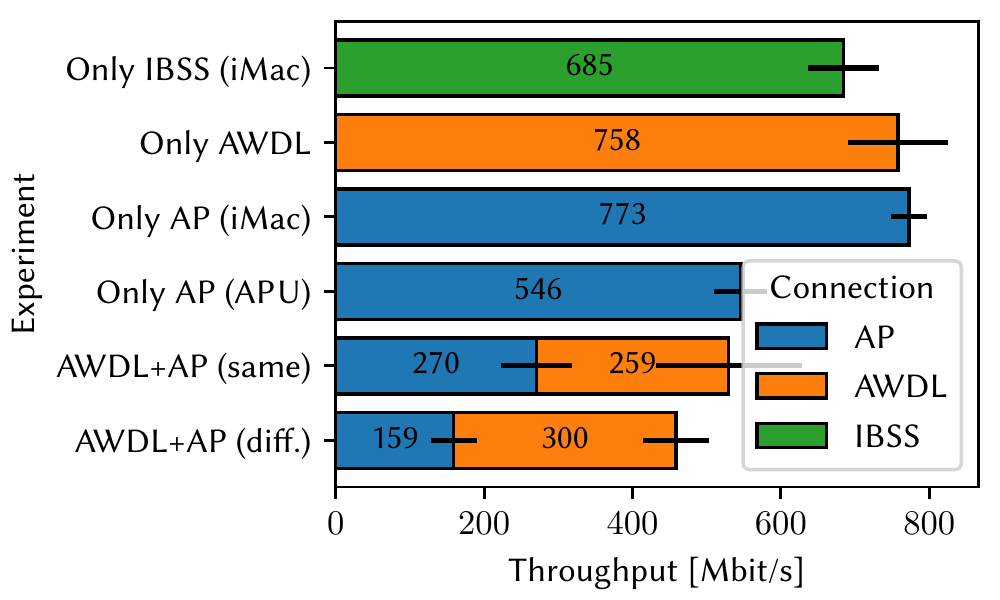}
	\caption{Throughput measurements.}
	\label{fig:iperf}
\end{figure}

\extended{
We already know that \gls{AWDL} makes use of highest possible PHY data rates for unicast frames if both participants support them and if the signal strength is high enough.
}{}
We want to evaluate the impact of \gls{AWDL}'s channel hopping on the throughput of a TCP connection.
Unfortunately, Apple drops packets for regular TCP and UDP servers that directly bind to the \code{awdl0} interface. This meant that running measurement software such as \code{iperf} was not immediately possible.
As a solution, we built an \gls{AWDL}--TCP proxy via the \code{NSNetService} API~\cite{Apple:NSNetServiceRef} which whitelists the advertised port. In essence, the proxy server advertises a service via DNS-SD and listens for incoming TCP connections. The proxy client component connects to it. Both proxy endpoints also allow TCP connections via the loopback interface such that regular TCP services can simply connect to the loopback interface, and forward the TCP traffic via the \code{NSNetService} connection.
The proxy tool is available at~\cite{proxawdl}.

\para{TCP Throughput}
We measure the throughput with \code{iperf} using three different nodes (MacBook, iMac, and an \gls{AP}) in six different settings:
(1) a single connection from MacBook to the \gls{AP} \emph{without} \gls{AWDL};
(2) a single connection from MacBook to iMac via \gls{AWDL} \emph{without} the \gls{AP};
(3) two concurrent connections as a combination of (1) and (2), while the \gls{AP} operates on channel 44; and
(4) as (3) but the \gls{AP} operates on channel 36.
Our sniffer is configured as an \gls{AP} in this scenario which supports a maximum PHY data rate of 866.7\,Mbit/s (MCS 9, two streams, 80\,MHz bandwidth). iMac and MacBook both support three streams. Thus, we include another measurement (5) where the iMac acts as the \gls{AP} to see possible throughput differences between an \gls{AWDL} and an \gls{AP} connection using the same hardware.
Finally, we include (6) a comparison to IEEE\,802.11 IBSS mode.
We repeat each 10-second experiment 50 times for each setting and show the results in \cref{fig:iperf}. The error bars indicate the standard deviation.
The \emph{only AWDL} and \emph{only AP (iMac)} settings result in similar throughput demonstrating that bandwidth is only limited by the hardware capabilities of the communicating nodes.
Note that the \emph{only IBSS (iMac)} setting performs 10--12\,\% worse than the previous two settings: we observed that the MCS selection mechanism for IBSS on macOS is erratic and does not always choose the maximum supported values even when the signal-to-noise ratio is high.
The Qualcomm Wi-Fi chips in the APU only support two streams, so the maximum bandwidth is reduced by approximately 30\,\%.
The cumulative throughput when the \gls{AP} operates on channel 44 (\emph{same}) is similar to the throughput of the \emph{only AP} setting while the bandwidth between the two connections is uniformly distributed.
When the \gls{AP} operates on a \emph{different} channel, the cumulative throughput drops by about 13\,\%. This confirms the intuitive assumption that channel switching affects throughput negatively. We are surprised to see that the bandwidth is no longer uniformly distributed between the two streams. Instead, \gls{AWDL} has a higher throughput which could be caused by \gls{AWDL} resorting to using all three available streams.

\para{Channel Hopping}
We found that \gls{AWDL} adopts its channel sequence according to the traffic volume on the interface.
When there is no traffic (such as in the \emph{idle} scenario), \gls{AWDL} allocates at least 25\,\% of the channel sequence to the social channels (see slots 1, 9, 10, and 11 in \cref{fig:activechan}). As the load increases, \gls{AWDL} may allocate all \glspl{EAW} for itself. We depict the various channel allocation states in \cref{tab:state-channellist}.\footnote{We found references for 25 states in total (including a \emph{real-time} mode and different combinations) during binary analysis, which we will not further discuss in this paper.}
The table shows that
\begin{inparaenum}[(1)]
	\item at least 25\,\% of the time is allocated for \gls{AWDL} (\emph{low power} state),
	\item there is \emph{always} a switch to channel 6 in slot 9 possibly for backward compatibility, and
	\item at least 25\,\% of the time is reserved for the \gls{AP} connection if the node is connected to an \gls{AP}.
\end{inparaenum}
In our throughput experiment, either the \emph{data} or the \emph{data+infra} (50\,\%) state was active.

\begin{table}
\caption{A subset of \gls{AWDL} states and corresponding channel list where $p$ and $s$ are the primary (44) and secondary (6) \gls{AWDL} channels, respectively, and $i$ is the channel of the \gls{AP}.}
\small
\label{tab:state-channellist}
\newcommand{\MyTabSpace}{\enspace}
\begin{tabular}{@{}lrc@{\MyTabSpace}c@{\MyTabSpace}c@{\MyTabSpace}c@{\MyTabSpace}c@{\MyTabSpace}c@{\MyTabSpace}c@{\MyTabSpace}c@{\MyTabSpace}c@{\MyTabSpace}c@{\MyTabSpace}c@{\MyTabSpace}c@{\MyTabSpace}c@{\MyTabSpace}c@{\MyTabSpace}c@{\MyTabSpace}c@{}}
	\toprule
	\textsc{State} & \textsc{Airtime} & \multicolumn{16}{@{}c@{}}{\textsc{Channel List ($c = 16$)}} \\
	\midrule
	Low Power & 25.0\,\% & $p$ & & & & & & & & $s$ & $p$ & $p$ & & & & & \\
	Idle & 37.5\,\% & $p$ & $p$ & $p$ & & & & & & $s$ & $p$ & $p$ & & & & & \\
	\multirow{2}{*}{Data+Infra} & 50.0\,\% & $p$ & $p$ & $p$ & $p$ & $i$ & $i$ & $i$ & $i$ & $s$ & $p$ & $p$ & $p$ & $i$ & $i$ & $i$ & $i$ \\
	& 75.0\,\% & $p$ & $p$ & $p$ & $p$ & $p$ & $p$ & $i$ & $i$ & $s$ & $p$ & $p$ & $p$ & $p$ & $p$ & $i$ & $i$ \\
	Data & 100.0\,\% & $p$ & $p$ & $p$ & $p$ & $p$ & $p$ & $p$ & $p$ & $s$ & $p$ & $p$ & $p$ & $p$ & $p$ & $p$ & $p$ \\
	\bottomrule
\end{tabular}
\end{table}

%% file: chapter/discussion.tex
\section{Discussion}
\label{sec:discussion}

In this section, we discuss \gls{AWDL} complexity and overhead, energy efficiency, and conduct an initial security assessment of \gls{AWDL} and its OS integration.

\subsection{Complexity and Overhead}

\gls{AWDL} has a complex protocol definition that supports various configurations using \glspl{AW} and \glspl{EW}. We were surprised to see that Apple settled for a static and rather simple configuration, making the complex concepts obsolete. In addition, we found a lot of redundant information that bloats the size of the \gls{AWDL} \glspl{AF}.

\begin{figure}
\centering
\includegraphics[width=\linewidth]{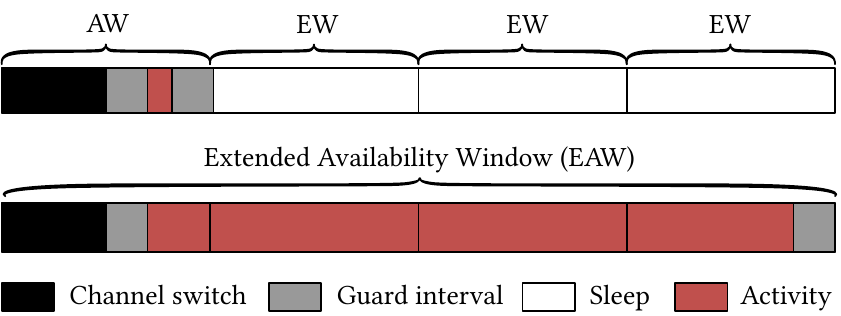}
\caption{Time spent for channel switching, guard interval, and resulting airtime that can be used for communication when using \glspl{AW}/\glspl{EW} vs.\ \glspl{EAW}.}
\label{fig:chan-switch-problem}
\end{figure}

\para{(Extended) Availability Windows}
\gls{AWDL}, as implemented in current OSes, allows for highly configurable operation configurations (see Synchronization Parameters \gls{TLV} in \cref{fig:syncparams}). 
However, all current implementations use a fixed channel sequence length of 16 and do not differentiate between \glspl{AW} and \glspl{EW} but exclusively use the longer \glspl{EAW} (compare \cref{fig:sync-sequence}).
The reason why Apple prefers \glspl{EAW} might have to do with the time that is required to perform a channel switch operation in the Wi-Fi chip. We found that a channel switch operation takes at least 8\,ms ($\approx$ 8\,\gls{TU}) using \code{dump chanswitch} of the \code{wl} utility. In combination with a guard interval that is necessary to cope with the accepted synchronization error of 3\,\gls{TU}, this would leave only 2\,\gls{TU} airtime for communication assuming that the \glspl{EW} are reserved for an energy conserving \emph{sleep} state. When using \glspl{EAW}, the temporal efficiency increases from about \,12.5\,\% to more than 78\,\% while sacrificing opportunities to save energy.
We visualize this difference in \cref{fig:chan-switch-problem}.

\para{Redundancy}
\gls{AWDL} \glspl{AF} contain redundant information such as the current master address which is announced in the Synchronization Parameters, Election Parameters, and Election Parameters v2 \glspl{TLV}.
The Service Response Parameters \gls{TLV} often encode the same information multiple times such that the service instance string and device name be seen three times in a frame when AirDrop is active.

\subsection{Energy Efficiency}

Our working hypothesis was that energy efficiency was one of the primary design goals of \gls{AWDL} (compare \cref{sec:overview}). The insights obtained from our experimental analysis do not support this hypothesis.
We have found that even in the so-called \emph{low power} state, \gls{AWDL} is active for at least 25\,\% of the time during which the Wi-Fi chip is active.
In addition, all nodes and not only the master send \glspl{PSF}.
We suspect that energy efficiency was sacrificed for a more reliable operation: the exclusive use of long \glspl{EAW} makes the system more robust against synchronization error. As all nodes send \glspl{PSF}, new nodes can discover an existing \gls{AWDL} cluster faster.

\subsection{Security}

\gls{AWDL} connections are completely unsecured. However, Apple employs a default packet filter that prevents services to listen on the \gls{AWDL} interface accidentally.
We further found and reported a vulnerability in the macOS driver interface.

\para{Open \gls{AWDL} Connection}
We have found that \gls{AWDL} connections do not feature any security mechanism.
All action and data frames are sent in plain and without authentication.
\gls{AWDL} delegates security functions to the transport and application layer, e.\,g., AirDrop uses TLS~1.2~\cite{Apple:iOSSecurity}.
The approach appears to be an informed decision to implement application-dependant policies:
a device might be trusted for sending an image file via AirDrop, but not for remote-controlling a Keynote presentation.

\para{Default Packet Filter}
While an \gls{AWDL} connection can be considered insecure, Apple made sure that other services such as file sharing are \emph{not} advertised via the \code{awdl0} interface which would otherwise be accessible by unauthenticated nearby adversaries.
Developers need to explicitly use a dedicated API (e.\,g., \code{NSNetService}) to opt-in for the use of \gls{AWDL} which we did to implement our TCP proxy.
The packet filter is apparently not part of the standard macOS firewall but probably implemented in \code{NSNetService}.
Also, the \code{awdl0} interface is activated only on demand and deactivated once no more traffic is registered, thus, minimizing the time window for an attack. This could be considered an ``accidental'' security mechanism because the main reason for the timeout was probably energy conservation.

\para{Vulnerable Driver Interface}
The \code{ioctl} interface described in \cref{sec:binary-analysis}, especially including the card-specific command used for the Broadcom \code{wl} utility, could be used by \emph{any local user} on macOS.
\begin{anonsuppress}
The issue was reported to Apple on July 19, 2017, and was assigned CVE-2017-13886. 
Apple has fixed this issue on December 6, 2017, and published the CVE entry on May 2, 2018~\cite{CVE-2017-13886}. 
\end{anonsuppress}


%% file: chapter/conclusion.tex

\section{Conclusion}
\label{sec:conclusion}

We reconstructed the frame format and the operation of \gls{AWDL}, a complex undocumented protocol and complemented our findings with an open source Wireshark dissector.
We believe that public knowledge of such wide-spread proprietary protocols is vital to assist wireless network operators and to allow independent security audits as well as to stimulate innovation and research below the application layer.
We experimentally evaluated \gls{AWDL} and showed that the synchronization accuracy is about -0.45\,ms on average. The maximum achievable throughput is only limited by the devices' supported PHY data rates if the nodes are not actively using an infrastructure network. If channel switching is required, the cumulative throughput of two concurrent connections drops by about 13\,\%.
We have found a security bug which allowed any local user to access the macOS Wi-Fi driver interface. In the light of recent over-the-air exploitable IEEE\,802.11 implementations~\cite{Beniamini2017:BroadcomPt2}, we suspect that there are even more vulnerabilities to be found given the complexity of the \gls{AWDL} protocol.
As future work, we will direct our efforts towards an energy model for \gls{AWDL} to understand the implications when using \gls{AWDL} as a drop-in replacement for \gls{BLE} or IEEE\,802.11 IBSS in ad hoc communication applications.